\documentstyle[prl,aps,epsf,multicol]{revtex}
\tighten
\renewcommand{\narrowtext}
{\begin{multicols}{2}
\global\columnwidth20.5pc}
\renewcommand{\widetext} 
{\end{multicols}\global\columnwidth42.5pc}
\begin{document} 
\draft 
\title{Dynamical scaling at the quantum Hall transition:\\ 
  Coulomb blockade versus phase breaking}
\author{D. G. Polyakov$^{1,*}$ and K. V. Samokhin$^{2,\dagger}$}
\address{$^{(1)}$ Institut f\"ur Theoretische Physik, Universit\"at zu
  K\"oln, Z\"ulpicher Str.\ 77, 50937 K\"oln, Germany\\ $^{(2)}$
  Cavendish Laboratory, University of Cambridge, Madingley Road,
  Cambridge CB3 0HE, UK} 
\maketitle 
\begin{abstract} 
We argue that the finite temperature dynamics of the integer quantum
Hall system is governed by two independent length scales. The
consistent scaling description of the transition makes crucial use of
two temperature critical exponents, reflecting the interplay between
charging effects and interaction-induced dephasing. Experimental
implications of the two-scale picture are discussed.
\end{abstract} 
\narrowtext

Scaling treatment of the Anderson metal-to-insulator transition is
central to understanding of the integer quantum Hall (QH) effect
\cite{huckestein95}. The plateau transitions are understood as
isolated critical points separating two localized phases, so that the
localization length $\xi$ only diverges at a discrete set of the
critical energies $E_c$. While a reliable analytical theory is sorely
missing, the scaling ideas have long served to correlate the results
of experiment and of numerical simulation. The observed {\it
dynamical} scaling, however, still presents a {\it puzzle} which has
defied a convincing explanation for almost a decade, starting from the
very first experiments \cite{wei88}. On the experimental side, the
scaling has been probed by tuning through the transition at different
temperatures (by varying the Landau level filling factor) and
observing how fast the critical singularities are rounded off with
increasing $T$. The experimental data tell us that the long-distance
cutoff $L_h$ scales as $T^{-1/z}$ with the dynamical critical exponent
$z=1$. Specifically, the dissipative {\it dc} conductivity $g$ (in
units of $e^2/h$) has the scaling form $g=g_cF(L_h/\xi)$, where
$F(0)=1$, $F(\infty)=0$, and $g_c\sim 1$. The traditional use of $z$
in this context is related to the common belief \cite{sondhi96} that
at criticality the only relevant temporal scale is $\tau\sim T^{-1}$.

It can be readily seen, however, that despite the simplicity of this
experimental picture, it implies the {\it inadequacy}, in describing
the QH critical point, of the usual theoretical framework
\cite{sondhi96} based on the assumption that the system at criticality
can be characterized by just one temporal scale $T^{-1}$. Indeed, the
peculiarity of the Anderson transition in two dimensions -- the
non-vanishing $g_c$ -- means that the QH system at the critical point
is {\it diffusive}, so that the irreducible dynamical susceptibility
is a function of $\omega/q^z$ with $z=2$ \cite{chalker88}. It follows
that if there are only two scales ($L_h$ and $\tau\propto L_h^z$) at
play, they must be related via the diffusion law ($z=2$). It has
become customary to refer to the Coulomb interaction between electrons
as the source of the ``anomalous" $z=1$. However, the long wave-length
diffusion coefficient $D=h^{-1}g_c/(\partial n/\partial\mu)$ is finite
in the interacting QH system as well, since for disordered electrons
the thermodynamic density of states (DOS) $\partial n/\partial\mu$
does not exhibit any singular behavior when the Coulomb interaction is
turned on, and we assume that the critical conductivity $g_c$ also
remains finite \cite{nb}. Likewise, the screening properties of the
integer QH metal can be described in terms of the usual RPA
response. In fact, the {\it only} peculiarity of the QH metallic
phase, as compared to a weakly disordered conventional metal, is a
fractal dispersion of the diffusion coefficient at large $q^2/\omega$
\cite{chalker88}. Thus the attempt to explain the cutoff $L_h\propto
\tau^{1/z}$ by introducing $\tau\sim T^{-1}$ and setting $z=1$
\cite{sondhi96} is confronted by the fact that electron dynamics at
the critical point is diffusive $(z=2)$.

Another recent attempt to substantiate the observed dynamical scaling
relates \cite{lee96} the apparent degradation $z=2\to z=1$ to the
linear vanishing of the one-particle DOS $\rho_1(\omega)\propto
|\omega|$ at the Fermi level $(\omega=0)$. This fault with dimension
counting underlines the common misconception of the problem once
more. First, it is misleading to insert the one-particle DOS in the
renormalization group machinery in place of $\partial
n/\partial\mu$. Moreover, there is every reason to question the very
assumption that $\rho_1\propto|\omega|$ at the {\it metallic} critical
point. We argue below that in actual fact $\rho_1(\omega)$ vanishes at
the QH transition faster than any power of $\omega$.

Apart from the purely scaling arguments, there is controversy about
the physical mechanism of the cutoff. Again, if one follows
\cite{sondhi96} and identifies the cutoff with the interaction-induced
dephasing length $L_\phi$, one encounters the difficulty in trying to
connect the $T^{-1}$ behavior of $L_h$ with the usual dependence
$L_\phi\propto T^{-1/2}$, which merely reflects the diffusive
character of transport of interacting particles and should be valid at
the QH critical point as well. Hence the concept \cite{sondhi96} of
the quantum-classical crossover controlled by the dephasing length
appears to be inadequate to the physics of the QH transition. Note,
however, that the discarding of $L_\phi$ is not quite trivial since
$L_\phi\ll L_h$ in the low-$T$ limit, which means that the {\it
shorter} of the two length scales is irrelevant.

In this paper, we attempt to sort out the problem of the dynamical
scaling. Our findings are as follows. The scaling description of the
integer QH transition for {\it interacting} electrons includes {\it
two} independent length scales, $L_h\propto T^{-1}$ and $L_\phi\propto
T^{-1/2}$. They govern the temperature driven scaling outwards and
towards the unstable fixed point \cite{khmelnitskii83}, respectively
(Fig.\ 1). Both are related to the corresponding temporal scales
$\tau_h$ and $\tau_\phi$ via the diffusion law ($z=2$): $\tau_h\sim
DL_h^2\propto T^{-2}$ and $\tau_\phi\sim DL_\phi^2\propto T^{-1}$. The
Coulomb interaction therefore does {\it not} change the {\it true}
dynamical exponent $z$ from 2 to 1; instead, it leads to the emergence
of the two different scales. It is only if one uses the usual
representation of the length scales in the form $L_h\propto
T^{-1/z_1}$ and $L_\phi\propto T^{-1/z_2}$ that there appears the
dynamical exponent $z_1=1$, whereas $z_2$ remains equal to 2
\cite{belitz94}. The typical energy transfer is $T$ and the
phase-breaking rate $\tau_\phi^{-1}$ is also of order $T$; however,
the scattering rate $\tau_h^{-1}$ behaves as $T^2$. The corresponding
cutoff $L_h$ has nothing to do with the phase breaking: the
temperature smearing of the transition is controlled by charging
effects similar to those in the Coulomb blockade regime. The shape of
the Coulomb gap in the one-particle DOS at the critical point has no
direct relation to either of the dynamical exponents $z_1$ or
$z_2$. Separately, we argue that $\rho_1(\omega)$ vanishes as
$\exp[-\alpha\ln^2(T_c/|\omega|)]$, where $\alpha\sim 1$ and $T_c$ is
a characteristic width of the gap. \begin{figure} \epsfxsize=2.0truein
\centerline{\epsfbox{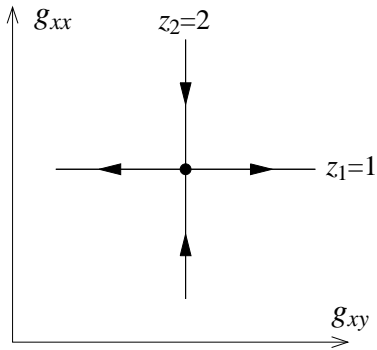}} \vspace{3mm}\caption{Scaling with
lowering $T$ outwards and towards the unstable fixed point is governed
by different length scales with temperature exponents $z_1=1$ and
$z_2=2$, respectively.}  \end{figure}

Our basic point in the description of the dynamical scaling is that
the QH system at the critical point is metallic (in contrast to the
critical system at a conventional Anderson transition in three
dimensions) and it makes perfect sense to treat it as an ordinary
dirty metal with $g\sim 1$. We therefore begin with the effect of
electron-electron scattering on the quantum interference of diffusons
\cite{hikami81} in a {\it weakly} disordered metal ($g\gg 1$) with
completely broken time-reversal symmetry. To the best of our
knowledge, this has not been spelled out clearly in the literature.
The diffusion propagator ${\cal D}_{\omega q}^{\omega_0}$ for
interacting electrons is a function of two frequencies -- only in the
absence of interactions ${\cal D}_{\omega q}^{\omega_0}\propto \delta
(\omega_0)$. It is convenient to choose the mixed representation ${\rm
D}_{\omega q}^{t_0}=\int {d\omega_0\over 2\pi}\exp
(-i\omega_0t_0){\cal D}_{\omega
  q}^{\omega_0}$ and regard the delay time $t_0$ as a parameter. The
Dyson's equation assumes then the algebraic form $[{\rm D}_{\omega
  q}^{t_0}]^{-1}=[{\rm D}_{\omega q}^{(0)}]^{-1}-\Sigma_{\omega
  q}^{t_0}$, where the bare propagator ${\rm D}_{\omega
  q}^{(0)}=1/(-i\omega + Dq^2)$. We define the {\it diffuson decay
  rate} $1/\tau_\phi^{\rm D}(t_0)=-{\rm Re} \Sigma_{\omega q}^{t_0}$
as a function of $t_0$ (assuming that the weak interaction does not
renormalize ${\rm D}_{\omega q}^{t_0}$ on the microscopic scale).
Particle number conservation dictates that $1/\tau_\phi^{\rm D}(0)=0$,
since the dynamical part of the density-density correlator $\left<
  nn\right>_{\omega q}$ is expressed in terms of the integral $\int
{d\omega_0\over 2\pi}{\cal D}_{\omega q}^{\omega_0}$. Thus, in
contrast to the more familiar Cooperon, ${\cal D}_{\omega
q}^{\omega_0}$ cannot be characterized by a {\it single}
phase-breaking time (this should also be contrasted with the cutoff of
the full diffusion propagator by a constant $\tau_\phi^{\rm D}$, cf.\
\cite{castellani86}). To calculate $1/\tau_\phi^{\rm D}(t_0)$, we use
the method \cite{altshuler82}, within the framework of which the
electron-electron interaction is mediated by thermal fluctuations of a
classical ($\omega\ll T$) electromagnetic field with the correlator
$\left<VV\right>_{\omega q}=4\pi e^2v_s
T/\varepsilon(\omega^2+v_s^2q^2)$, where $v_s=(e^2/\varepsilon\hbar)g$
is the charge-spreading velocity, $\varepsilon$ the bare dielectric
constant (Nyquist noise). We transform to real space by writing the
equation for the diffuson in the form \begin{eqnarray}
&&\left[{\partial\over \partial t} - D{\partial^2\over\partial{\bf
r}^2}\right.\hfill \\ &&\left.  +{i\over\hbar}\left(V({\bf
r},t-{t_0\over 2}) - V({\bf r},t+{t_0\over 2})\right)\right]{\rm
D}^{t_0}({\bf r},t)=\delta({\bf r})\delta(t)~.\nonumber \end{eqnarray}
Notice the crucial difference between this equation and that for the
Cooperon (cf.\ \cite{altshuler82}): in the latter case the times $t$
and $t_0$ are interchanged in the argument of the effective potential;
as a result, $t_0$ becomes a ``mute variable", -- the averaged
Cooperon does not depend on $t_0$ and this is why it is characterized
by the single time $\tau_\phi^{\rm C}$. Calculating the correlator of
the potential in Eq.\ (1), we observe that $\tau_\phi^{\rm D}(t_0)$
can be obtained similarly to $\tau_\phi^{\rm
  C}$ by introducing the effective interaction
$\left<VV\right>_{\omega k}^{t_0} =\left<VV\right>_{\omega
  k}(1-\cos\omega t_0)$. It follows immediately that in the limit
$t_0\gg\tau^{\rm D}_\phi(t_0)$, where the oscillating term $\cos\omega
t_0$ can be safely ignored, the particle-hole and particle-particle
propagators decay in the same way: $\tau_\phi^{\rm
  D}(\infty)=\tau_\phi^{\rm C}$. The difference shows up at smaller
$t_0$: one gets with logarithmic accuracy the equation for the decay
rate of ${\rm D}^{t_0}_{\omega q}$: $1/\tau_\phi^{\rm
  D}(t_0)=2\int\!\!{d^2{\bf k}\over (2\pi)^2}\int\!\!{d\omega\over
  2\pi}\left<VV \right>^{t_0}_{\omega,{\bf k}+{\bf q}}{\rm Re}{\rm
  D}^{t_0}_{\omega k}$. Solving it, we obtain the compact expression
\begin{equation} {1\over\tau_\phi^{\rm D}(t_0)}={T\over g}\ln {T\over
D\max \{q^2, (Dt_0)^{-1}, [v_s\tau_\phi^{\rm
D}(t_0)]^{-2}\}}~. \end{equation} This formula tells us that for
$q\sim [D\tau_\phi^{\rm D}(t_0)]^{-1/2}$, which are relevant in the
calculation of the conductivity, the decay rate starts to fall off as
$\ln (Tt_0)$ at $t_0\alt\tau_\phi^{\rm D}(\infty)$. In the extreme of
small $t_0\ll T^{-1}$ the quasiclassical treatment is no longer
accurate, but an estimate can be readily obtained by cutting off the
frequency integration at $\omega\sim T$, -- it follows that the
dephasing rate vanishes algebraically at zero $t_0$:
${1/\tau_\phi^{\rm D}(t_0)}\sim(T/g)(T t_0)^2$.

Now let us look at the effect of the interaction on the quantum
interference of diffusons. In the unitary limit, the leading
weak-localization correction is given by the familiar expression
$\delta g^{\rm D}\sim g^{-1}\ln (L/l)$ \cite{hikami81}, where $l$ is
the mean free path (or the Larmor radius, when it is smaller), $L$ an
inelastic scattering length. However, the mechanism of the infrared
cutoff in the high-$B$ limit deserves comment, since the dephasing
time $\tau_\phi^{\rm D}(t_0)$ tends to infinity as $t_0\to 0$. The
quasiclassical treatment of the Coulomb interaction allows to
calculate first the contribution to $g$ from diffusons ${\rm
D}^{t_0}({\bf r},t)$ moving in a given (as if externally applied)
Nyquist potential. The Gaussian average over the thermal
electromagnetic fluctuations ($\left<\ldots\right>$ below) can then be
safely performed. For the leading correction, this gives $\delta
g^{\rm D}=g^{-1}\int_0^{\infty}\!\!  dt\left<A(t)\right>$, where
$A=A_2+A_3$ is a sum of two- and three-diffuson terms \cite{hikami81}
(a proper cutoff on the ballistic scale is assumed).  Consider the
simplest two-diffuson contribution \begin{eqnarray} A_2(t)=
2D^2\int_0^{t}\!\!dt'{\rm D}^{t'-t}(0,t') {\rm D}^{t'}(0,t-t')~,
\end{eqnarray} which already reveals the peculiarity of the dephasing
in the unitary case. Though one could have expected that
$\left<A_2(t)\right>$ would decay exponentially at $t\gg\tau_\phi^{\rm
D}(\infty)$, it can be readily seen from Eq.\ (3) that
$\left<A_2(t)\right>$ remains singular on the scale of $\tau_\phi^{\rm
D}(\infty)$. The phase coherence is preserved because of the vanishing
of the dephasing rate at $t'=0$ and $t'=t$. A similar ``breakdown" of
the dephasing occurs in $\left<A_3(t)\right>$. However, adding all the
pieces, we find that the total contribution to $\delta g^{\rm D}$,
$\left<A(t)\right>\propto\exp [-t/\tau^{\rm D}_\phi(\infty)]$, decays
on the scale of the shortest dephasing time. This proves that the
interaction-induced cutoff for $\delta g^{\rm D}$ is given by the
phase-breaking length related to $\tau^{\rm D}_\phi(\infty)$ (which
contrasts with the result of Ref.\ 12, where the inelastic cutoff of
the weak localization in the unitary limit was identified with a much
longer energy-relaxation length).

We turn now to the interaction-induced dephasing at the integer QH
transition. We assume that the interaction is weak enough not to break
down the integer QH effect, i.e.\ $e^2/\varepsilon\lambda\ll\Gamma$,
where $\lambda$ is the magnetic length, $\Gamma$ the width of the
disorder-broadened Landau level. It is then legitimate to repeat the
above analysis of the phase breaking right at the QH metallic point by
endowing the diffusion coefficient with a strong dispersion at
$Dq^2/\omega\agt 1$ \cite{chalker88}. The power-law dispersion at
large $q^2/\omega$ signals that the QH metal starts to develop the
critical eigenfunction correlations.  However, as follows from the
calculation with constant $D$, this does not change the dependence of
$L_\phi$ on $T$, since the relevant $Dq^2/\omega$ are of order
unity. Specifically, an estimate can be readily obtained by setting
$g\sim 1$ in Eq.\ (2), which gives $T\tau_\phi^{\rm D}(\infty)\sim 1$
and $L_\phi\sim (D/T)^{1/2}$ ($z_2=2$). Notice that when the Fermi
energy coincides with $E_c$, the localization effects can be neglected
at all $\omega\ll\Gamma$, since $\xi\gg (D/\omega)^{1/2}$ within the
energy band of width $\omega$ around $E_c$. In sum, the scale on which
the dephasing occurs at the critical point is certainly
$(D/T)^{1/2}\ll L_h$. We are led to conclude that while the phase
breaking controls the temperature scaling of $g_c$ right at the
critical point, it does {\it not} control the observed metal-insulator
crossover.

The reason for the strong increase of the cutoff $L_h$ as compared to
$L_\phi$ is that away from the critical point transport is governed by
charging effects: the Coulomb blockade on the scale of $\xi$
drastically narrows the crossover region. Indeed, one can identify two
characteristic energies on the scale of $\xi$: the charging energy
$U_c\sim e^2/\varepsilon\xi$ and the ``on-site" energy spacing
$\Delta\sim 1/(\partial n/\partial\mu)\xi^2$. Near the transition
$U_c\gg\Delta$. The naive description of scaling in terms of
$L_\phi/\xi$ amounts to the assumption that the QH system shows
crossover at $T/\Delta\sim 1$. It is evident, however, that the system
behaves as a metal only if $T$ exceeds $U_c$, -- otherwise the
scattering is blocked as in the usual Coulomb blockade regime. The QH
system at given $E_F$ can thus be modeled as a dense array of quantum
dots of size $\xi$ coupled via the tunneling integral $\sim
\Delta$. The scaling form of $g$ then reads \begin{equation}
g=g_cF(U_c/T)~, \end{equation} or, equivalently, $g=g_cF(L_h/\xi)$
with $L_h\sim e^2/\varepsilon T$, so that $z_1=1$ (these arguments
parallel those in \cite{polyakov93}, where $F(x)$ was argued to fall
off at $x\to\infty$ as $\ln F\sim -x^{1/2}$). Hence, the scaling
around the unstable fixed point indeed necessitates dealing with {\it
two} scales, $L_h$ and $L_\phi$ (Fig.\ 1). Also, while the typical
energy transfer and the dephasing rate are both $\sim T$, the
scattering rate $\tau_h^{-1}\sim DL_h^{-2}$ is much smaller:
\begin{equation} 1/\tau_h\sim T^2/T_c~,\quad T_c\sim
e^4/\varepsilon^2D~.  \end{equation}

To test the two-scale picture with $z_1\neq z_2$ experimentally, we
suggest to measure the temperature dependent correction to the
critical conductivity $\delta g_c(T)$. Specifically, according to
numerical simulations \cite{huckestein95,wang96}, the finite-size
correction to $g_c$ scales as $L^{-y}$ with $y\simeq 0.4\div 0.5$ (in
fact, it can be shown analytically \cite{janssen97} that $y$ is not an
independent exponent, namely there exists the non-trivial relation
$y=\eta$, where $\eta\simeq 0.4$ is the usual critical exponent of
eigenfunction correlations \cite{chalker88}). We predict that, while
the smearing of the transition is controlled by $L_h$ ($z_1=1$), the
critical conductivity scales with $L_\phi$ ($z_2=2$), i.e.\ $\delta
g_c\propto T^{y/2}$. Another possible test is based on the fact that
$L_\phi\ll L_h$. Naively, one may well think that when $L_h$ becomes
larger, as $T\to 0$, than the system size $L$, there must appear
strong mesoscopic fluctuations (say of the height of the conductivity
peak). However, our approach suggests that this is not true, since in
the range $L_\phi\ll L\ll L_h$ the width of the critical region is
already $T$ independent but the mesoscopic fluctuations are still
suppressed (at $\tau_\phi^{-1}\sim T$, the only parameter that governs
the amplitude of the fluctuations is $L^2T/D$). The absence of the
fluctuations at $L_h\agt L$ would give a strong experimental support
to the two-scale picture.

Finally, we discuss briefly the behavior of the one-particle DOS at
the critical point $\rho_{1c}(\omega)$. It is a popular misconception
that the reduction $z_1\to 1$ signifies the linear vanishing of
$\rho_{1c}(\omega)\propto |\omega|$ (see, e.g., \cite{lee96}). In
fact, several aspects require comment. First, as argued above, the
true dynamical exponent is related to $g_c$ and the {\it
thermodynamic} DOS $\partial n/\partial\mu$, so that it is equal to 2
at the QH transition (merely reflecting the Einstein relation).
Second, away from the critical point, the quasiparticle DOS $\rho_h$
that appears in the hopping exponent \cite{polyakov93} indeed behaves
as $\rho_h\sim |\omega|\varepsilon^2/e^4$ at $|\omega|\alt U_c$;
however, $\rho_h$ does {\it not} coincide with $\rho_1$ unless the
system is classical and electrons can be treated as point charges.
The difference is due to the fact that in the classical treatment of
the Coulomb gap \cite{efros85} $\Delta/|\omega|$ is sent to $\infty$,
whereas near the critical point $\Delta$ is the smallest energy
scale. As a result, the rate of the charge spreading becomes a crucial
factor in the suppression of $\rho_{1c}$ in the metallic phase. The
width of the interaction-induced gap in a metal grows with decreasing
$g$ as $\exp [-2(\pi g)^{1/2}]$ \cite{altshuler85}. To calculate
$\rho_{1c}(\omega)$, we use the elegant quasiclassical method
suggested in \cite{levitov97}, which works well in the conducting
phase even if $g\sim 1$. Adjusting it to the high-$B$ limit (in our
case the screening length $D/v_s$ is larger than the Larmor radius),
we obtain at $g_c\sim 1$ \begin{equation} \rho_{1c}(\omega)= (\partial
n/\partial\mu)\exp [-S(\omega)]~,\, S\simeq
\alpha\ln^2(T_c/|\omega|)~, \end{equation} where the numerical
coefficient $\alpha\sim 1$, and the width of the gap $T_c$ is defined
by Eq.\ (5). It is worth noticing that the localization-induced
dispersion of the diffusion coefficient at large $q^2/\omega$, which
is the only peculiarity of the QH critical point as compared to the
Drude metal, is of little importance here (in contrast to the
conventional two-dimensional metal, where the localization effects get
in the way of the method \cite{levitov97} at $\omega\to 0$). Note also
the shape of the gap at the transition, -- $\rho_{1c}$ vanishes faster
than any power of $\omega$. This should be contrasted with both the
power-law behavior of $\rho_1$ at the Anderson transition in
$2+\epsilon$ dimensions \cite{finkelstein84} and the naive power
counting at the QH transition \cite{lee96}. This result also brings up
the question whether the Hartree-Fock method \cite{ericyang93}, within
the framework of which a linear vanishing of $\rho_{1c}$ was observed
numerically, captures all the essential physics. Away from the
critical point, the ``log-normal" suppression of the DOS saturates
with decreasing $\omega$ at $|\omega|\sim U_c$ (which means that the
charge spreading stops on the scale of $\xi$). In the insulating
phase, the linear \cite{efros85} vanishing of $\rho_1$ should be
expected at $\omega=0$, but with a slope suppressed by the factor of
$\exp[-S(U_c)]$.

To summarize, we have argued that the temperature driven scaling at
the integer QH transition is governed by two independent length scales
with the temperature exponents $z_1=1$ and $z_2=2$. The smearing of
the transition is controlled by charging effects ($z_1=1$), whereas
the interaction-induced phase breaking $(z_2=2)$ is responsible for
corrections to the critical conductivity. We suggested experimental
tests of the two-scale picture.

We thank J. Hajdu, B. Halperin, D.~Khmelnitskii, A.~Mirlin, N.~Read,
and P.~W\"olfle for interesting discussions. This work was supported
by the Deutsche Forschungsgemeinschaft, by the German-Israeli
Foundation, and by the Russian Foundation for Basic Research under
grant No.\ 96-02-17894a.

\end{multicols}

\begin{references}
\bibitem{nb} The scaling form of the irreducible density-density
correlator ${\rm Im}\left<nn\right>_{\omega q}=q^{2-z}f(\omega/q^z)$
is inconsistent, unless $z=2$, with the condition that both $\partial
n/\partial\mu =\lim_{q\to 0}\left<nn\right>_{0q}$ and
$g_c=-ih\lim_{\omega\to 0}\omega\lim_{q\to
0}q^{-2}\left<nn\right>_{\omega q}$ are finite.
\end{references}
\end{document}